\documentclass{article}

\usepackage{latexsym}
\usepackage{epsf,axodraw,graphicx}

\def\sl{\llap{$/$}}

\begin{document}

\title{\bf Factorization and Unitarity in Superstring Theory}

\author{
Zhi-Guang Xiao\footnote{E-mail: xiaozhiguang@pku.edu.cn or
zgxiao@itp.ac.cn; Present address: Interdisciplinary Center of
Theoretical Studies, Chinese Academy of Sciences, P. O. Box 2735,
Beijing 100080, P. R. China}
\\[2mm]\\
School of Physics, Peking University \\
Beijing 100871, P. R. China\\ \\
Chuan-Jie Zhu\footnote{E-mail: zhucj@itp.ac.cn}
\thanks{Supported
in part by fund from the National Natural Science Foundation of
China with grant Number 10475104.} \\[2mm]\\
Institute of Theoretical Physics,
Chinese Academy of Sciences\\
P. O. Box 2735, Beijing 100080,
P. R. China\footnote{Permanent address}\\[2mm]\\
and   \\[2mm]\\
 Abdus Salam International Center for Theoretical Physics\\
Strada Costiera 11, I-34014 Trieste, Italy}

\maketitle

\abstract{The overall coefficient of the two-loop 4-particle
amplitude in superstring theory is determined by making use of the
factorization and unitarity. To accomplish this we computed in
detail all the relevant tree and one-loop amplitudes involved and
determined their overall coefficients in a consistent way. }

\newpage
%%%%%%%%%%%%%%%%%%%%%%%%%%%%%%%%%%%%%%%%%%%%%%%%%%%%%%%%%%%%%%%%%%

\section{Introduction}

Explicit result for higher loop amplitudes in superstring is quite
rare. To our knowledge the only explicitly known higher loop ($\ge
2$) non-vanishing amplitude is the four-particle amplitude in
superstring theory, firstly obtained in \cite{IengoZhu} and later
re-obtained in \cite{WuZhu1,WuZhu2} in an explicitly gauge
independent way, following the works of D'~Hoker and Phong
\cite{DHokerPhonga,DHokerPhongb,DHokerPhongc,DHokerPhongd,
DHokerPhongdd, DHokerPhonge} on two loop measure of superstring
theory. This result was also computed in a super-Poincare
covariant way in \cite{Nathan}. Recently D'~Hoker and Phong
\cite{DHokerPhongaa,DHokerPhongbb} also gave a measure for three
loop superstring theory. It  remains to see if this can be used to
do explicit three loop computations in superstring theory. For
another promising approach of covariant calculation of superstring
amplitudes we refer the reader to Berkovits' review
\cite{Berkovits}.

Due to the rareness of explicit results, it is natural to study
the known result in depth. The old result was cast into an
explicit modular invariant form \cite{IengoZhu2} and used in
\cite{Iengo} to prove the vanishing of the $R^4$ correction
\cite{GrossWitten,Green}. It has also been proved in \cite{WuZhu3}
that the results obtained in \cite{IengoZhu,WuZhu1,WuZhu2} are
equivalent. Another goal we have in mind is to make connection
with known results from field theory in ${\cal N}=4$
supersymmetric Yang-Mills theory \cite{Berna,Bernb}. It seems
natural to compute the precise overall coefficient of the two-loop
4-particle amplitude.

This seems a trivial problem, but in fact it turns out to be quite
involved. One could try to use factorization and unitarity to fix
the overall coefficient for the four-particle amplitude. In a
previous paper \cite{XiaoZhu} we have studied in detail the
factorization of the two-loop 4-particle amplitude in superstring
theory\footnote{See also \cite{Zhu,Yasuda} for early works on
two-loop factorization.}. When we use this result to determine the
coefficient of the two-loop 4-particle amplitude, we found that we
need  the precise overall factor for other one-loop amplitudes
involved. Due to the incomplete results in literature (and a fear
of wrongly quoting other's results), we therefore computed all the
relevant amplitudes in a consistent way and fixed all the overall
factors by either using factorization or unitarity.  In this way
the coefficient of the 2-loop 4-particle amplitude is  determined
exactly. This paper is organized as follows:

In the next section we recall all the vertex operators needed in
this paper, following the covariant quantization of the
Neveu-Schwarz-Ramond theory by Friedan, Martinec and Shenker
\cite{FMS} and Knizhnik \cite{Knizhnik}. In section 3 we gave all
the results for tree amplitudes needed with their overall
coefficients. We omit all the computations as this becomes
standard exercises in superstring theory. We don't claim any
originality for these amplitudes although some amplitudes with
massive tensor may be new. The one-loop amplitudes are collected
in section 4. Starting from the 4-particle amplitude we obtained
the 3-particle (one massive tensor) and 2-particle (both massive
tensor) amplitudes by factorization. The overall coefficient is
determined by using unitarity (which is the only place we used
unitarity relation). The summation over the intermediate states is
quite involved and 2 appendices  are devoted to the proof of an
equation (eq.~(\ref{eqappendixc})). In section 5 we combine the
results of the previous sections and the result of \cite{XiaoZhu}
to determined the overall coefficient of the 2-loop 4-particle
amplitude.

During the writing of this paper we received the paper of D'Hoker,
Gutperle and Phong \cite{Phong3} which also determined the precise
overall coefficient of the 2-loop 4-particle amplitude. The method
used for the final determination of the 2-loop coefficient is the
same. Nevertheless they used the result of S-duality for the
determination of the 1-loop coefficient. We found complete
agreement with their result although the factorization is
performed by using hyperelliptic language.

\section{A review of the vertex operators}

First let us set our notations for the vertex operators used. For
all our calculations we use the covariant emission vertices
constructed by Friedan, Martinec and Shenker \cite{FMS}, and by
Knizhnik \cite{Knizhnik}. Let us briefly review their
construction. Most of the time we will present the result for the
left-moving (or holomorphic) part, as the formalism for the
right-moving (or anti-holomorphic) part is the same. So it is
useful to introduce a set of notations to separate the two parts.
As we will use only three different vertex operators in this
paper, we use the subscripts $B$ and $F$ to denote the  bosonic
and fermionic massless vertex operators from the Neveu-Schwarz
sector and Ramond sector respectively. The level one (or the
first) massive vertex operator from the Neveu-Schwarz sector is
denoted by a subscript $M$. Subscripts with an additional  \~ \,
denote the right-moving part when we need the complete amplitude
in superstring theory.

By using these notations, the vertex operator for the massless
$NS$-$\widetilde{NS}$ tensor is:
\begin{equation}
 \mathcal{V}_{B\widetilde{B}}^{(-1,-1)}
 (z,\bar z, k, \epsilon, \tilde\epsilon)
 = \mathcal{V}_{B}^{(-1)}(z,k, \epsilon )
\mathcal{V}_{\widetilde{B}}^{(-1)}(\bar z,k, \tilde\epsilon ),
\end{equation}
where
\begin{eqnarray}
 \mathcal{V}_{B}^{(-1)}(z,k, \epsilon ) & = &
 g_c\, \epsilon\cdot\psi(z) \, {\rm e}^{-\phi(z)} \,
  {\rm e}^{ik\cdot X(z, \bar z)} ,
 \label{leftparta} \\
\mathcal{V}_{\widetilde{B}}^{(-1)}(\bar z,k, \tilde\epsilon ) & =
& \tilde{\epsilon}\cdot\tilde{\psi}(\bar{z}) \,
 {\rm e}^{-\tilde{\phi}(\bar{z})} \, .
\end{eqnarray}
Here in the above, we have written the polarization tensor in a
factorized form: $\epsilon_{\mu\nu} =
\epsilon_\mu\tilde\epsilon_\nu$. By convention we   absorbed the
overall constant $g_c$ and the exponential factor ${\rm
e}^{ik\cdot X(z, \bar z)} $ into the left-moving vertex operator
$\mathcal{V}_{B}$.  One may also split $X(z,\bar z)$ into a
left-moving part and a right-moving part, but we will not do this
in this paper as this is not essential for our purpose.

We note that the vertex operator given in eq.~(\ref{leftparta})
carries a ghost charge of $-1$. We will also need the physical
equivalent vertex operator which carries no ghost charge. It is
given as follows:
\begin{eqnarray}
\mathcal{V}_{B}^{(0)}(z,k,\epsilon) & = & -g_c(
\epsilon\cdot\partial{X}(z) + i k\cdot\psi(z){\epsilon}
\cdot\psi(z)  )
 \, {\rm e}^{ik\cdot X(z)} \, .
\end{eqnarray}
This vertex operator is obtained from the ghost charge $-1$
operator of eq.~(\ref{leftparta}) by using the ``picture-raising"
operator $Z(y)$:
\begin{eqnarray}
Z(y) & = & \{Q, 2 \xi(y) \} =  - P\cdot \psi \, {\rm e}^\phi +
\cdots ,
\\
\mathcal{V}_{B}^{(0)}(z,k,\epsilon) & = &   :Z(z)
\mathcal{V}_{B}^{(-1)}(z,k,\epsilon): \nonumber \\
&  = & {1\over 2 \pi i } \oint_z { {\rm d}  y  \over y - z} \,
Z(y) \mathcal{V}_{B}^{(-1)}(z,k,\epsilon) ,
\end{eqnarray}
modulo spurious operators\footnote{Exactly we have:
$$
:Z(z) \psi^\mu(z)\, {\rm e}^{-\phi(z) + i k \cdot X(z)} :=
 ( i k^\mu (\eta \xi + \partial \phi) - (\partial X^\mu
+ i k\cdot \psi \psi^\mu) ) \, {\rm e}^{i k \cdot X(z)} .
\nonumber
$$}. All other vertex operators of different ghost charge can be
related in the same way. They are physically equivalent
\cite{FMS}.

The second vertex operator (left-moving part only) is the massive
tensor from the Neveu-Schwarz sector
\cite{Iengoa,MassiveVertexa,MassiveVertexb}:
\begin{eqnarray}
\mathcal{V}_{_M}^{(-1)}  & = & g_{_M} \, \Big\{
\alpha_{\mu\nu\rho} i\psi^\mu(z)\psi^\nu(z) \psi^\rho(z) \nonumber
\\ & &  + \sigma_{\mu\nu}
\partial{X}^\mu(z)\psi^\nu(z) - \sigma_\mu i\partial{\psi}^\mu(z)
\Big\}\, {\rm e}^{-\phi(z)}  \, {\rm e}^{ik\cdot X(z)} \, ,
\end{eqnarray}
which is in the $(-1)$-picture (or ghost charge $-1$) and
\begin{eqnarray}
{\cal V}^{(0)}_{_M}
  & = & -g_{_M}\Big\{\alpha_{[\mu\nu\rho]}\big[3 i\partial
 X^\mu\psi^\nu\psi^\rho + {\alpha' \over 2}\,  k\cdot\psi
 \psi^\mu\psi^\nu\psi^\rho\big] \nonumber \\
 & & + \left(2\over \alpha'\right)^{1/2} \,
  \sigma_{\mu\nu} \big[ \partial X^\mu\partial
 X^\nu+ {\alpha' \over 2} \,
 (\partial \psi^\mu \psi^\nu+ik\cdot\psi \partial X^\mu
 \psi^\nu) \big] \nonumber \\
 & & -\sigma_\mu\big[
 i\partial^2 X- {\alpha' \over 2} \, k\cdot\psi \partial
 \psi^\mu\big] \Big\} \, {\rm e}^{ik\cdot X} ,
\end{eqnarray}
which is in the $0$-picture and the dependence on the dimensional
scale $\alpha'$ is restored for ${\cal V}^{(0)}_{_M}$. The
mass-shell condition is $k^2 = - { 4 \over \alpha'}$. Here
$\alpha_{\mu\nu\rho}$ and $\sigma_{\mu\nu}$ are the polarization
tensors which satisfies the following normalization conditions:
\begin{equation}
\alpha_{\mu\nu\rho}(k)\alpha^{\mu\nu\rho}(-k) = -{1\over 6},
\qquad \sigma_{\mu\nu}(k)\sigma^{\mu\nu}(-k) = 1 .
\end{equation}
which are given in \cite{XiaoZhu}. The state represented by
$\sigma_\mu$ is null and it will not appear in the physical
amplitude.

The last vertex operator is the massless fermion from the Ramond
sector:
\begin{equation}
 V_F^{(-\frac{1}{2})}   =  g_{_F} {\rm
e}^{-\phi/2}u^\alpha S_\alpha
 \, {\rm e}^{-\tilde{\phi}(\bar{z})}\, {\rm e}^{ik\cdot X(z)},
\end{equation}
which is in the $(-{1\over2})$-picture and $u$ is a Majorana-Weyl
spinor in 10-dimensional space-time. We will not need the
expression in the ${1\over2}$-picture which we give it here
\begin{equation}
 V_F^{( \frac{1}{2})}   =  g_{_F} u^\alpha\Big\{ {\rm e}^{\phi/2}
 [ \partial X^\mu + {i\over 4}\, k\cdot \psi \psi^\mu]
 \gamma_{\mu \alpha\beta} \, S^\beta + {1\over 2}\,
 {\rm e}^{3\phi/2}\eta b S_\alpha\Big\} {\rm e}^{i k \cdot X},
\end{equation}
just for completeness (see \cite{FMS} for details).

As a last note, our convention for the $S$-matrix is:
\begin{equation}
S(1,\cdots, N) = \delta(1,\cdots,N) + (2\pi)^D\delta^{D}(k_1 +
\cdots + k_N) \, i {\cal A}_N(1,\cdots, N) ,
\end{equation}
where all momenta are incoming and $D=10$ for superstring theory.
All the formulas are given in terms of ${\cal A}_N$ and the
momentum conservation is implicit in it.

\section{The tree-level amplitudes and their factorization
in superstring theory}

\subsection{The massless boson amplitudes}

A general tree-level $n$-particle (assuming to be all massless NS
bosons) amplitude is computed as follows:
\begin{eqnarray} i{\cal A}_{n}(k_i, \epsilon_i)
& = & \int \prod_{i=4}^n {\rm d}^2z_{i}
 \Big\langle
 \big[c \mathcal{V}_{B}^{(-1 )}\big](z_1,k_1,\epsilon_1)
 \big[c \mathcal{V}_{B}^{(0 )}\big](z_2,k_2,\epsilon_2)
\nonumber \\
& & \times
 \big[c \mathcal{V}_{B}^{(-1 )}\big](z_3,k_3,\epsilon_3)
\prod_{i=4}^n
 \big[c \mathcal{V}_{B}^{(0)}\big](z_i,k_i,\epsilon_i) \nonumber
 \\
& &  \times (\hbox{right-moving part}) \Big\rangle,
 \end{eqnarray}
by fixing the first three insertion points of the vertex operators
and integrating the rest insertion points \cite{Polchinski}. To
obtain a non-trivial amplitude ($n\ge4$) we do need the
right-moving part explicitly. The computation is straightforward
but quite tedious.

The results for $n=3$ and $n=4$ are well-known and are given as
follows \cite{Polchinski}:
\begin{eqnarray}
 {\cal A}_3(k_i,\epsilon_i,\tilde{\epsilon}_i)
 &  = & { 8\, \pi  \, g_c \over \alpha'}   \, K_{3}(k_i,\epsilon_i)
K_{3}(k_i,\tilde \epsilon_i) , \\
 {\cal A}_{4 }(k_i,\epsilon_i, \tilde{\epsilon}_i)
 & = & c\times {-  \kappa^2 (\alpha')^3 \over 4}
 \, K(k_i,\epsilon_i)
 {K} (k_i,\tilde\epsilon_i) \nonumber \\
& & \times   {\Gamma(-{\alpha' s\over 4} )\Gamma(-{\alpha' t\over
4})\Gamma(-{\alpha'u\over 4}) \over \Gamma(1+{\alpha'
s\over4})\Gamma(1+{\alpha' t\over4})\Gamma(1+{\alpha' u\over4})} ,
\label{eqoneone}
\end{eqnarray}
where the various kinematic factors are given as follows:
\begin{eqnarray}
K_{3}(k_i,\epsilon_i) & = & ( \alpha'/ 2)^{1/2} ( \epsilon_1\cdot
\epsilon_2 \epsilon_3\cdot k_1 +\epsilon_1\cdot
\epsilon_3\epsilon_2\cdot
k_3+\epsilon_2\cdot\epsilon_3 \epsilon_1\cdot k_2 ) , \nonumber \\
 K (k_i,\epsilon_i) & = & -{(\alpha')^2 \over 16}\Big{[} ut
 \epsilon_1\cdot\epsilon_2\epsilon_3\cdot\epsilon_4+st
 \epsilon_1\cdot\epsilon_3\epsilon_2\cdot\epsilon_4+us
 \epsilon_1\cdot\epsilon_4\epsilon_2\cdot\epsilon_3\Big]\nonumber\\
 & &
 +{t(\alpha')^2 \over 8}\Big[
 \epsilon_1\cdot\epsilon_2\epsilon_3\cdot
 k_1\epsilon_4\cdot k_2+\epsilon_1\cdot\epsilon_3\epsilon_2\cdot
 k_1\epsilon_4\cdot k_3 \nonumber \\
 & & \qquad +\epsilon_2\cdot\epsilon_4\epsilon_1\cdot
 k_2\epsilon_3\cdot k_4+\epsilon_3\cdot\epsilon_4\epsilon_2\cdot
 k_4\epsilon_1\cdot k_3\Big] \nonumber\\& &
 +{u(\alpha')^2\over 8}\Big[ \epsilon_1\cdot\epsilon_2\epsilon_3\cdot
 k_2\epsilon_4\cdot k_1+\epsilon_1\cdot\epsilon_4\epsilon_2\cdot
 k_1\epsilon_3\cdot k_4 \nonumber \\
 & & \qquad +\epsilon_2\cdot\epsilon_3\epsilon_1\cdot
 k_2\epsilon_4\cdot k_3+\epsilon_3\cdot\epsilon_4\epsilon_1\cdot
 k_4\epsilon_2\cdot k_3\Big]\nonumber\\& &
 +{s(\alpha')^2\over 8}\Big[ \epsilon_1\cdot\epsilon_3\epsilon_2\cdot
 k_3\epsilon_4\cdot k_1+\epsilon_2\cdot\epsilon_4\epsilon_1\cdot
 k_4\epsilon_3\cdot k_2 \nonumber \\
 & & \qquad +\epsilon_2\cdot\epsilon_3\epsilon_1\cdot
 k_3\epsilon_4\cdot k_2+\epsilon_1\cdot\epsilon_4\epsilon_2\cdot
 k_4\epsilon_3\cdot k_1\Big] ,
\end{eqnarray}
which are dimensionless. Here $K(k_i,\epsilon_i)$ is the standard
(left-moving part) kinematic factor from the tree, one-loop and
two-loop computations in superstring theory
\cite{GreenSchwarz,IengoZhu,WuZhu2}. In our convention all the
kinematic factors are dimensionless (and they agree with the
previous ones after setting $\alpha'=2$). The right-moving part
kinematic factors   are obtained from the corresponding
left-moving part kinematic factors by the simple substitution
$\epsilon_i \to \tilde\epsilon_i$. For example we have:
\begin{equation}
 K_{3 }(k_i,\tilde\epsilon_i) = ( \alpha'/ 2)^{1/2}(
 \tilde\epsilon_1\cdot
\tilde\epsilon_2 \tilde\epsilon_3\cdot k_1 + \tilde\epsilon_1\cdot
\tilde\epsilon_3\tilde\epsilon_2\cdot k_3 +\tilde\epsilon_2
\cdot\tilde\epsilon_3 \tilde\epsilon_1\cdot k_2 )  .
\end{equation}
This rule will apply for all later formulas involving right-moving
part contribution if we don't explicitly say an alternative.

The constant $c$ appearing in (\ref{eqoneone}) can be proved to be
equal $1$ by using factorization property of the amplitude. In
order to do this, we must use the following result for the
summation over intermediate states:
 \begin{eqnarray}
  & &  \hskip -2cm
  \sum_{\epsilon({k})}  K_{3}(k_1,\epsilon_1; k_2,\epsilon_2;
 k, \epsilon(k)) \,
 K_{3}(k_3, \epsilon_3; k_4,\epsilon_4;
 -k, \epsilon(-k))   \nonumber \\
  & = & - { 4 \over u\alpha' } \, K(k_i, \epsilon_i)|_{s=0, t = -u} ,
\end{eqnarray}
which can be easily proved by using the following formula:
\begin{equation}
\sum_{\epsilon(k) } \epsilon_\mu(k) \,\epsilon_\nu(-k) =
\sum_{\epsilon(k) } \epsilon_\mu(k) \,\epsilon^*_\nu(k)=
\eta_{\mu\nu} - {1\over 2  k\cdot p} \, ({k_\mu p_\nu + k_\nu
p_\mu}) ,
\end{equation}
where $p$ is a reference momentum and can be chosen as $(p^\mu) =
(k^0, - k^i)$.

\subsection{The two massless boson and one massive tensor vertex}

The left-moving part vertex for two massless boson and one massive
tensor is:
\begin{eqnarray}
i \mathcal{A}_{2BM} & = & \Big\langle
 \big[c \mathcal{V}_{B}^{(-1 )}\big](z_1,k_1)
 \big[c \mathcal{V}_{B}^{(0 )}\big](z_2,k_2)
 \big[c \mathcal{V}_{M}^{(-1 )}\big](z,k)
\Big\rangle
\nonumber \\
 & = & i g_{_M}  \, K_{M}(k_1,\epsilon_1; k_2, \epsilon_2;
 k, \alpha, \sigma),
\end{eqnarray}
where the kinematic factor $K_{M}$ is:
\begin{eqnarray}
K_{M} & = &  -6\, \left(\alpha'\over 2\right)^{1/2}
\alpha_{\mu\nu\rho}
 k_1^\mu\epsilon^\nu_1\epsilon_2^\rho+  \sigma_{\mu\nu}
 \big[ \epsilon_1^\mu\epsilon_2^\nu \nonumber \\
 & & + {\alpha'\over 2} \, ( k_1^\mu k_1^\nu
 \epsilon_1\cdot\epsilon_2
 -k_1^\mu\epsilon_1^\nu
 \epsilon_2\cdot k_1+ k_1^\mu\epsilon_2^\nu \epsilon_1\cdot
 k_2 ) \big]  . \label{eqtwoone}
 \end{eqnarray}
As we said before, there is no contribution from the $\sigma_\mu$
term of the vertex operator as it gives spurious physical states.

By combining the left-moving part vertex with right-moving part
vertex, we can use the factorization property of the 4-particle
amplitude for $s \to {4\over \alpha'}$ to obtain the overall
coefficient $g_{_M}$ as we did in the last subsection. Here we
must use the following formula for the summation over intermediate
massive states:
\begin{eqnarray}
& &  \hskip -1.5cm \sum_{\alpha(k), \sigma(k)}
K_{M}(k_1,\epsilon_1; k_2, \epsilon_2;
 k, \alpha(k), \sigma(k) ) \nonumber \\
 & &  \times  K_{M}(k_3,\epsilon_3; k_4, \epsilon_4;
-k, \alpha(-k), \sigma(-k)) = K(k_i, \epsilon_i)|_{s = {4\over
\alpha'}} . \label{eqproofa}
\end{eqnarray}
This is proved in \cite{XiaoZhu}.

\begin{figure}[htb]
\begin{center}\includegraphics[width=10cm]{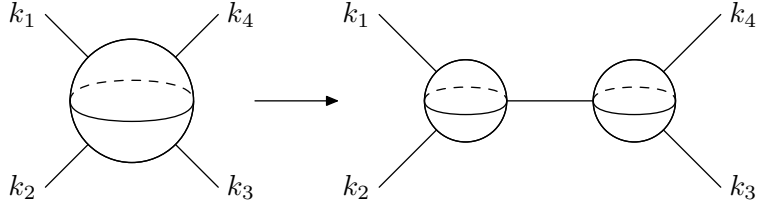}\end{center}
\caption{The factorization of the 4-particle tree amplitude. The
intermediate state is a massive tensor.}
\end{figure}

By using eq.~(\ref{eqproofa}) and the factorization of the
4-particle amplitude into two 3-particle (one is massive)
amplitude as shown in Fig.~1, we get
\begin{equation}
g_{_M} = {4 \kappa \over \alpha'} = {8 \pi g_c \over \alpha'} .
\end{equation}
by choosing it to be positive.

\subsection{Tree amplitudes with fermions or R-R tensors}

In this subsection we only list all the required results. They are
needed to prove eq.~(\ref{eqappendixc}). It can be skipped if one
is only interested in the result for the overall coefficient.

First we have two 3-particle vertices. These are given as follows:
\begin{itemize}
\item Vertex $(NS,\widetilde{NS}) \to
(R,\widetilde{NS})+(R,\widetilde{NS})$:
\begin{equation}
{\mathcal A}_{F}(k_1,u_1,\tilde\epsilon_1;
k_2,u_2,\tilde\epsilon_2; k_3, \epsilon_3, \tilde\epsilon_3) = -
g_{_F} \, \sqrt{{\alpha'\over 2}}\bar{u}_1 \epsilon \sl_3 u_2 \,
K_{3}(k_i, \tilde \epsilon_i) ,
\end{equation}
with
\begin{equation}
 g_{_{F}} = { 4 \, \pi \, g_c \over \alpha'} .
\end{equation}

\item Vertex $(NS,\widetilde{NS}) +(R,\widetilde{R})
\to(R,\widetilde{R})$:
\begin{eqnarray}
\mathcal{A}_{_{R}}(k_1,u_1,\tilde u_1; k_2,u_2,\tilde u_2; k_3,
\epsilon_3, \tilde\epsilon_3) & = &
g_{_{R}}K_{_{R}}\tilde{K}_{_{R}}
\end{eqnarray}
where
\begin{eqnarray}
g_{_{R}} & = & {2 \pi \, g_c \over \alpha'} ,  \\
K_{_{R}} & = &  \sqrt{{\alpha'\over 2}}\, \bar{u}_1 \Gamma^\mu
u_2{ \epsilon_3}_\mu  .
\end{eqnarray}
\end{itemize}

For the 4-particle amplitude, the massless 2 fermion and 2 boson
($(R,\widetilde{NS}) +(NS,\widetilde{NS})
\to(R,\widetilde{NS})+(NS,\widetilde{NS})$) amplitude is:
\begin{equation}
{\mathcal A}_{FBFB} = g_{_{2F}} \, K_{FBFB} \, \tilde{K}(k_i,
\tilde\epsilon_i) \, {\Gamma(-{\alpha' s\over 4} )\Gamma(-{\alpha'
t\over 4})\Gamma(-{\alpha'u\over 4}) \over \Gamma(1+{\alpha'
s\over4})\Gamma(1+{\alpha' t\over4})\Gamma(1+{\alpha' u\over4})} ,
\end{equation}
with
\begin{equation}\
g_{_{2F}} =  { 2 \pi^2 \,  g_c^2  \over  \alpha' } .
\end{equation}
Here the kinematic factor $K_{FBFB}$ is:
\begin{eqnarray}
K_{FBFB} &=&  - {t(\alpha')^2\over 4}\, \Big[ \epsilon_4\cdot k_3
\bar{u_1}\epsilon\sl_2u_3+\frac{1}{2}{k_4}_\mu {\epsilon_4}_\nu
\bar{u}_1\epsilon\sl_2\Gamma^{[\mu\nu]}u_3\Big] \nonumber \\
& & -{s(\alpha')^2\over 4}\, \Big[\epsilon_4\cdot k_1
\bar{u_1}\epsilon\sl_2u_3+\frac{1}{2}{k_4}_\mu {\epsilon_4}_\nu
\bar{u}_1\Gamma^{[\mu\nu]}\epsilon\sl_2u_3 \Big] .
\end{eqnarray}

The  amplitude  with 2 Ramond-Ramond tensors ($ (NS,
\widetilde{NS}) + (R,\widetilde{R}) \to(NS,\widetilde{NS})
+(R,\widetilde{R})$) is:
\begin{equation}
\mathcal{A}_{_{RBRB}} = -g_{_{2R}}\, K_{_{FBFB}} \,
\tilde{K}_{_{FBFB}} \,  {\Gamma(-{\alpha' s\over 4}
)\Gamma(-{\alpha' t\over 4})\Gamma(-{\alpha'u\over 4}) \over
\Gamma(1+{\alpha' s\over4})\Gamma(1+{\alpha'
t\over4})\Gamma(1+{\alpha' u\over4})} ,
\end{equation}
with
\begin{equation}
g_{_{2R}} =  {  \pi^2 \, g_c^2  \over \alpha'}  .
\end{equation}
The (right-moving) kinematic factor $\tilde{K}_{_{FBFB}}$ is
obtained from $K_{_{FBFB}}$ by putting a tilde on every $u$ and
$\epsilon$.

We also need two 3-particle vertices with one massive tensor. The
results are:
\begin{itemize}
\item  Massive boson of $(NS,\widetilde{NS})\to$ 2 massless
fermion $(R,\widetilde{NS})+(R,\widetilde{NS})$:
\begin{eqnarray}
& & \hskip -1.5cm {\cal A} _{MFF}(k,\alpha,\sigma,
\tilde\alpha,\tilde\sigma; k_1,u_1, \tilde\epsilon_1; k_2,
u_2,\tilde\epsilon_2) \nonumber \\
&  = &  g_{_{MFF}}K_{_{MFF}} (k,\alpha,\sigma; k_1,u_1; k_2, u_2)
\nonumber \\
& & \times  \tilde{K}_{M} (k,\tilde\alpha,\tilde\sigma;
k_1,\tilde\epsilon_1; k_2, \tilde\epsilon_2),   \\
g_{_{MFF}} &  = & { 4 \pi \, g_c \over \alpha'} ,
\end{eqnarray}
where the kinematic factor $K_{MFF}$ is:
\begin{equation}
K_{MFF}  =
 \frac{1}{2} \,\sqrt{\alpha'\over 2}  \alpha_{\mu\nu\rho}
 \bar{u}_1\Gamma^{\mu\nu\rho }u_2
- \frac{\alpha'}{2} \,  \sigma_{\mu\nu}k_2^\mu\,
\bar{u}_1\Gamma^\nu u_2  . \label{eqthreesix}
\end{equation}

\item  Massive boson $(NS,\widetilde{NS}) \to $ massless tensor
$(R,\widetilde{R})+(R,\widetilde{R})$:
\begin{equation}
\mathcal{A}_{MRR}  = -g_{_{MRR}}K_{_{MFF}}\tilde{K}_{_{MFF}},
\qquad  g_{_{MRR}} =  {2 \pi  \, g_c  \over \alpha'} .
\end{equation}
\end{itemize}

\section{One loop amplitudes and their factorization and unitarity}

\subsection{One loop amplitudes and their factorization}

At one loop, the four particle amplitude was  firstly computed by
Green and Schwarz in \cite{GreenSchwarz} and the result is:
\begin{eqnarray}
{\cal A}_4^{\rm 1-loop} & = &   g_4^{\rm 1-loop} \, K(k_i,
\epsilon_i) \, \int_F { {\rm d}^2 \tau \over ({\rm Im}\tau)^2 } \,
\int \prod_{i=1}^4 {{\rm d}^2 z_i \over {\rm Im}\tau } \,
\nonumber \\
& &  \times  \prod_{r<s}\left| {\Theta_1(z_{rs}|\tau) \over
\partial \Theta_1(0|\tau) } \, {\rm exp}\left(  {  -\pi \over {\rm
Im} \tau }({\rm Im}z_{rs})^2 \right) \right|^{\alpha' k_r\cdot
k_s} ,
\end{eqnarray}
where $z_{rs} = z_r - z_s$. What we required in this paper is the
3-particle amplitude with one massive tensor at 1-loop. We can
obtain it by factorization but we must make an assumption that the
kinematic factor is the same as that appearing at tree level. This
was explicitly checked by an explicit calculation at one-loop
\cite{Xiao}. Now we derive this 3-particle amplitude by using
factorization.

\begin{figure}[htb]
\begin{center}\includegraphics[width=10cm]{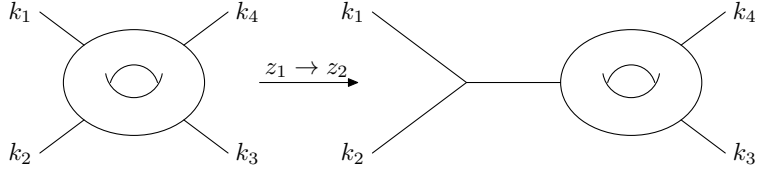}
\end{center}
\caption{The colliding limit of $z_1 \to z_2$ gives a 3-particle
one-loop amplitude.}
\end{figure}

By taking the limit of $z_1 \to z_2$ to select the physical pole
term as $s = -(k_1+k_2)^2 \to {4\over\alpha'}$, we have:
\begin{eqnarray}
{\cal A}_4^{\rm 1-loop} & \to &  g_4^{\rm 1-loop} \, K(k_i,
\epsilon_i) \, { - {4 \pi \over \alpha'} \over s - {4 \over
\alpha'}  } \int_F { {\rm d}^2 \tau \over ({\rm Im}\tau)^2 } \,
\int {{\rm d}^2 z_2 \over ({\rm Im} \tau)^2} \, \prod_{i=3}^4
{{\rm d}^2 z_i \over {\rm Im}\tau } \,
\nonumber \\
& & \hskip -1cm \times  \prod_{2\le r<s}\left|
{\Theta_1(z_{rs}|\tau) \over
\partial \Theta_1(0|\tau) } \, {\rm exp}\left(  {  -\pi \over {\rm
Im} \tau }({\rm Im}z_{rs})^2 \right) \right|^{\alpha' k_r'\cdot
k_s'} ,
\end{eqnarray}
where $k_2' = k_1 + k_2$ and $k_{3,4}' = k_{3,4}$. From this
result we can extract the (1-loop correction to the) 3-particle
amplitude by using the factorization limit as shown in Fig.~2. We
have:
\begin{eqnarray}
\mathcal{A}^{\rm 1-loop}_{2BM}(  k_1  ,k_2 ,k  ) & = & g_{3}^{\rm
1-loop} \,  K_{M} \tilde{K}_{M}  \int_F { {\rm d}^2 \tau \over
({\rm Im}\tau)^5 } \, \int \prod_{i=1}^2
{{\rm d}^2 z_i } \nonumber \\
& &  \hskip -1cm \times \prod_{   r<s}^3\left|
{\Theta_1(z_{rs}|\tau) \over
\partial \Theta_1(0|\tau) } \, {\rm exp}\left(  {  -\pi \over {\rm
Im} \tau }({\rm Im}z_{rs})^2 \right) \right|^{\alpha' k_r \cdot
k_s } , \label{eqfourty} \\
g_4^{\rm 1-loop} & = & { g_{_M} \, g_3^{\rm 1-loop} \over 4
\pi/\alpha' }    , \label{eqfourone}
\end{eqnarray}
by setting $k_3= k$. Here we have used the invariance of the
integrand under translation of all the insertion points to fix
$z=z_3$ to an arbitrary point on the torus (so there is no
integration over $z_3$ in eq.~(\ref{eqfourty})). Now we can use
factorization again to this amplitude by taking the limit $z_1 \to
z_2$ to select the physical pole term as $s = -(k_1+k_2)^2 \to
{4\over\alpha'}$. We have:
\begin{eqnarray}
\mathcal{A}^{\rm 1-loop}_{2BM} & \to &   g_3^{\rm 1-loop} \, K_{M}
\tilde{K}_{M}  \, { - {4 \pi \over \alpha'} \over s - {4 \over
\alpha'}  } \int_F { {\rm d}^2 \tau \over ({\rm Im}\tau)^5 } \,
\int {{\rm d}^2 z_2 }
\nonumber \\
& &   \times   \left| {\Theta_1(z_{2}|\tau) \over
\partial \Theta_1(0|\tau) } \, {\rm exp}\left(  {  -\pi \over {\rm
Im} \tau }({\rm Im}z_{2})^2 \right) \right|^{\alpha' (k_1+k_2)
\cdot k }  \, .
\end{eqnarray}
From this result we can extract the (1-loop correction to the)
2-particle amplitude by using the factorization limit as shown in
Fig.~3. We have:
\begin{eqnarray}
\mathcal{A}^{\rm 1-loop}_{MM}(  k, k')
 & = &   g_{_{MM}}^{\rm 1-loop} \,  K_{MM} \tilde{K}_{MM} \int_F {
{\rm d}^2 \tau \over ({\rm Im}\tau)^5 } \, \int {{\rm d}^2 z }
\nonumber \\
& &  \times   \left| {\Theta_1(z |\tau) \over
\partial \Theta_1(0|\tau) } \, {\rm exp}\left(  {  -\pi \over {\rm
Im} \tau }({\rm Im}z )^2 \right) \right|^{\alpha' k \cdot k'} ,
\\
g_3^{\rm 1-loop} & = & { g_{_M} \, g_{_{MM}}^{\rm 1-loop} \over {4
\pi \alpha'} } , \label{eqfourfour}
\end{eqnarray}
where the kinematic factor $K_{MM}$ is:
\begin{equation}
K_{MM}   =   -6{\alpha }_{\mu_1\nu_1\rho_1}(k)
{\alpha}^{\mu_1\nu_1\rho_1}(k')
+{\sigma}_{\mu\nu}(k){\sigma}^{\mu\nu}(k') .
\end{equation}

\begin{figure}[htb]
\begin{center}\includegraphics[width=10cm]{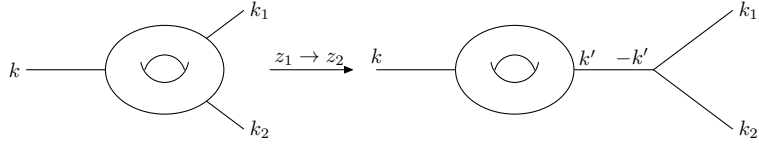}\end{center}
\caption{Further degeneration $z_1 \to z_2$ gives the 2-particle
one-loop amplitude.}
\end{figure}

\subsection{The one-loop unitarity relation for the massive tensor
propagator}

We will use the following formula from the operator formalism of
string theory:
\begin{eqnarray}
& & \hskip -1cm {\rm Tr}( V(k_1, {\rm e}^{2 \pi i z_1}) V(k_2,
{\rm e}^{2 \pi i z_2})\cdots V(k_M, {\rm e}^{2 \pi i z_M}) \,
q^N\, \bar q^{\tilde N} (q\bar q)^{ {\alpha'\over 4 }\hat p^2} )
\nonumber \\
& = & i \left[ -2 \pi \over \alpha' \ln|q| \right]^{D/2}
\prod_{r<s} \left| 2 \pi   {\Theta_1(z_{rs}|\tau) \over \partial
\Theta_1(0|\tau) } \, {\rm exp}\left(  { -\pi \over {\rm Im} \tau
}({\rm Im}z_{rs})^2 \right) \right|^{\alpha' k_r\cdot k_s} ,
\nonumber \\
\end{eqnarray}
where $q = {\rm e}^{2 \pi i z_M} = {\rm e}^{ 2 \pi i \tau} $ and
${\rm Im z_1}\le {\rm Im z_2} \le \cdots \le {\rm Im z_M} = {\rm
Im} \tau$. The vertex operator $V$ is:
\begin{eqnarray}
V(k, {\rm e}^{2 \pi i z} ) & = &  {\rm e}^{ \left(
\alpha'\over2\right)^{1/2} \sum_{n=1}^\infty {1\over n}\left(
k\cdot \alpha_{-n} \, {\rm e}^{ 2 \pi i n\, z} + k\cdot \tilde
\alpha_{-n} \, {\rm e}^{ 2 \pi i n\, \bar z} \right)} \nonumber \\
& & \times   {\rm e}^{ i k\cdot (\hat x + i 2 \pi\alpha' {\rm Im}
z \hat p)}  \nonumber \\
& & {\rm e}^{ - \left( \alpha'\over2\right)^{1/2}
\sum_{n=1}^\infty {1\over n}\left( k\cdot \alpha_{n} \, {\rm e}^{
-2 \pi i n\, z} + k\cdot \tilde \alpha_{n} \, {\rm e}^{- 2 \pi i
n\, \bar z} \right)} ,
\end{eqnarray}
and $V_R(k) = V(k, {\rm e}^{2 \pi i z} )|_{z \to 0, \hat x \to 0}$
which appears soon.

The trace contains an integration over momentum and summations
over all the Fock states created by the creation operators
$\alpha_{-n}$ and $\tilde\alpha_{-n}$ and their super-partners
(omitting the zero modes) (which are required to cancel the factor
$|\eta(\tau)|^D$ in bosonic string theory).

For $M=2$ after evaluating the trace by inserting a complete set
of intermediate states, we have
\begin{eqnarray}
& & \hskip -1cm { i (2 \pi)^{\alpha' k\cdot k'} \over (\alpha'
{\rm Im\tau})^{D/2} }  \left|   {\Theta_1(\tau - z|\tau) \over
\partial \Theta_1(0|\tau) } \, {\rm exp}\left(  { -\pi \over {\rm
Im} \tau }({\rm Im}(\tau - z))^2 \right) \right|^{\alpha' k \cdot
k'}  \nonumber \\
& = & \int{\rm d}^D p \left( {\rm e}^{-4\pi {\rm Im}z }\right)^{
{\alpha'\over 4} \, p^2} \, \left( {\rm e}^{-4\pi {\rm Im}(\tau -
z) }\right)^{ {\alpha'\over 4} \, (p+k')^2} \nonumber \\
& & \times \sum_{n, \tilde n, m, \tilde m} {\rm e}^{ 2 \pi i (n\,
z - \tilde n \bar z)} \, \langle n,\tilde n | V_R(k)|m , \tilde
m\rangle \nonumber \\
& & \hskip 1cm \times  \langle m,\tilde m | V_R(k')|n , \tilde
n\rangle \, {\rm e}^{ 2 \pi i (m\, (\tau -z) - \tilde m (\bar\tau
- \bar z))} .
\end{eqnarray}
Now we can do the integration over $\tau$ and $z$ explicitly. We
introduce an ultraviolet cutoff $\Lambda$ by restricting the
integration to the following region:
\begin{equation}
{\rm Im}\tau \ge 2 \Lambda , \qquad {\Lambda  } \le {\rm Im} z \le
{\rm Im}\tau - {\Lambda  } .
\end{equation}
After carrying out the integration explicitly we have
\begin{eqnarray}
& & \hskip -1.5cm { i (2 \pi)^{\alpha' k\cdot k'} \over (\alpha'
)^{D/2} } \int_{{\rm Im}\tau \ge \Lambda} { {\rm d}^2 \tau \over (
{\rm
Im\tau})^{D/2} } \int'{\rm d}^2 z \, \nonumber \\
&  & \times \left| {\Theta_1(\tau - z|\tau) \over
\partial \Theta_1(0|\tau) } \, {\rm exp}\left(  { -\pi \over {\rm
Im} \tau }({\rm Im}(\tau - z))^2 \right) \right|^{\alpha' k \cdot
k'}  \nonumber \\
& = & \int { {\rm d}^D p \over (\pi \alpha')^2} \sum_{n, \tilde n,
m, \tilde m}  { {\rm e}^{-\Lambda(p^2 + {4\over \alpha' } n - i
\epsilon) - \Lambda ((p-k)^2 + {4\over \alpha' } m - i \epsilon) }
 \over (p^2 + {4\over \alpha' } \, n - i \epsilon)
 ((p-k)^2 + {4\over \alpha' } m - i \epsilon) } \nonumber \\
 & & \hskip 1cm  \times \delta_{n,\tilde n}\delta_{m,\tilde m}\,
  \langle n,\tilde n | V_R(k)|m ,
\tilde m\rangle  \,  \langle m,\tilde m | V_R(k')|n , \tilde
n\rangle  .
\end{eqnarray}
By assuming $s = - k^2 = k\cdot k' \to {4 \over \alpha'}$, we see
that only the ``ground state'' could contribute an imaginary part
in the above summation (see below in eq.~(\ref{eqfiftytwo})), and
we have:
\begin{eqnarray}
i {\cal A}_{MM}^{\rm 1-loop} & = & g_{MM}^{\rm 1-loop} \,
K_{MM}\tilde K_{MM} \,  { (\alpha' )^{D/2}  \over (2 \pi)^{\alpha'
k\cdot k'}}
\nonumber \\
& & \times \int { {\rm d}^D p \over (\pi \alpha')^2}  \,
  { {\rm e}^{-\Lambda(p^2  + (p-k)^2 - i \epsilon) }
 \over (p^2   - i \epsilon)
 ((p-k)^2  - i \epsilon) }  \nonumber \\
 & & + i\,  (\hbox{Purely real parts}) .
\end{eqnarray}

From field theory or by explicit calculation, we have:
\begin{eqnarray} i A(s +  i \epsilon) &
= & \int{{\rm d}^D p \over (2 \pi )^D  } \, { 1\over p^2 + m^2 - i
\epsilon }\, { 1\over (p-k)^2 + m^2 - i \epsilon} \nonumber \\
{\rm Disc}A(s) & = & \Theta(s - 4 m^2)\, { i S_{D-1}\over
(2\pi)^{D-2}} \, {\left({s \over 4} -
m^2\right)^{D-3\over 2} \over 4 s^{1/2} } \nonumber \\
& = & i \, \int {{\rm d^D} k_1 \over (2\pi)^D } \, 2 \pi
\delta(k_1^2+m^2) \,\int {{\rm d^D} k_2
\over (2\pi)^D } \, 2\pi \delta(k_2^2+m_2^2) \, \nonumber \\
& & \times (2 \pi)^D\delta^D(k_1+k_2+k) \, |A^{\rm tree}(k_1; k_2;
k)|^2 . \label{eqfiftytwo}
\end{eqnarray} where
$s = - k^2$ and ${\rm Disc}A(s)\equiv A(s+i\epsilon) - A(s - i
\epsilon)$.

The unitarity relation is:
\begin{eqnarray} {\cal A}(s+i\epsilon)
- {\cal A}(s - i \epsilon) & = & { i \over 2!} \, \int {{\rm d^D}
k_1 \over (2\pi)^D } \, 2 \pi \delta(k_1^2) \,\int {{\rm d^D} k_2
\over (2\pi)^D } \, 2\pi \delta(k_2^2) \, \nonumber \\
& & \hskip -1cm \times (2 \pi)^D\delta^D(k_1+k_2+k) \, |{\cal
A}^{\rm tree}_{M**}(k_1; k_2; k)|^2 ,
\end{eqnarray}
where the factor of 2 is due to the propagation of identical
particles.

\begin{figure}[htb]
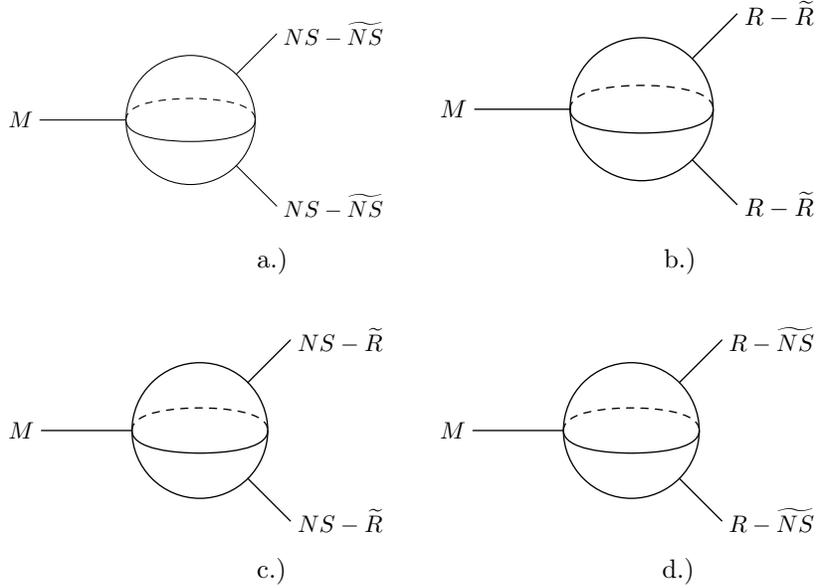

\begin{center}
\includegraphics[width=5cm]{threepoints.1} \hspace{.5cm}
\includegraphics[width=5cm]{threepoints.2}\end{center}
\hspace{4cm}\mbox{a.)}\hspace{5cm}\mbox{b.)} \vspace{.5cm}
\begin{center}
\includegraphics[width=5cm]{threepoints.3} \hspace{.5cm}
\includegraphics[width=5cm]{threepoints.4}\end{center}
\hspace{4cm}\mbox{c.)}\hspace{5cm}\mbox{d.)} \caption{The 4
possible contributions to ${\cal A}^{\rm tree}_{M**}$.}
\end{figure}

For superstring theory in ten dimensions, the possible ${\cal
A}^{\rm tree}_{M**}$'s are listed in Fig.~4. In Appendix we will
prove the following result for the summation over all possible
intermediate states:
\begin{equation} \sum_{\rm all~intermediate~states}
|{\cal A}_{M**}^{\rm tree}|^2 = (g_{_M})^2\, K_{MM}\tilde K_{MM}.
\label{eqappendixc} \end{equation} We relegate the proof of this
result to Appendix B.

By using this result we have:
\begin{equation}
g_{MM}^{\rm 1-loop} \,{ (\alpha' )^{D/2}  \over (2 \pi)^{\alpha'
k\cdot k'}} \, { (2 \pi)^D \over   (\pi \alpha')^2 } = {
(g_{_M})^2 \over 2}. \label{eqfivefive}
\end{equation}
by using this equation with eqs.~(\ref{eqfourone}) and
(\ref{eqfourfour}) we have:
\begin{eqnarray}
g_{_M}^{\rm 1-loop} & = & {g_c^2 \over 2 \pi^2 (\alpha')^5} , \\
g_{3}^{\rm 1-loop} & = & {g_c^3 \over   \pi^2 (\alpha')^5 }, \\
g_{4}^{\rm 1-loop} & = & {2 g_c^3 \over   \pi^2 (\alpha')^5}   .
\end{eqnarray}
The coefficient $g_{4}^{\rm 1-loop}$ agrees with the result of
Sakai and Tanni \cite{Sakai}.

\section{The factorization of the two loop four particle amplitude}

In this section we will use the result of \cite{XiaoZhu} to
determine the overall coefficient of the two-loop 4-particle
amplitude. We pay particular attention to the overall coefficient.
To begin with, let us recall the two-loop four-particle amplitude
in type II superstring theories obtained in
refs.~\cite{WuZhu1,WuZhu2}:
\begin{eqnarray}
\mathcal {A}_{II} & = &
  C_{II}  K(k_i,\epsilon_i) \, { 1\over 6!} \,
 \int {1\over T^5}
{\prod _{i=1}^6{\rm d}^2a_i\over {\rm d}
V_{pr}|\prod_{i<j}a_{ij}|^2} \nonumber \\
& & \times \prod_{i=1}^4 {{\rm d}^2z_i\over|y(z_i)|^2} \prod_{i<j}
 \exp\{-k_i\cdot k_j \langle X(z_i)X(z_j) \rangle
\},\nonumber \\
& & \times
\Big|s(z_1z_2+z_3z_4)+t(z_1z_4+z_2z_3)+u(z_1z_3+z_2z_4)\Big|^2 ,
\end{eqnarray}
where
\begin{eqnarray}
{\rm d} V_{pr} & = & {{\rm d}^2 a_i {\rm d}^2 a_j {\rm d}^2 a_k
\over
|a_{ij}a_{ik}a_{jk}|^2}, \\
 T & = & \int {\rm d}^2 z_1{\rm d}^2z_2
\frac{|z_1-z_2|^2}{|y(z_1)y(z_2)|^2}, \\
y^2(z) & = & \prod_{i=1}^6\, (z-a_i)\, ,
\end{eqnarray}
and $\langle X(z_i) X(z_j)\rangle\equiv \langle X(z_i,\bar{z}_i)
X(z_j,\bar{z}_j)\rangle$'s are the scalar correlators. The
$K(k_i,\epsilon_i)$ is the standard kinematic factor appearing at
tree, one- and two-loop computations
\cite{GreenSchwarz,IengoZhu,WuZhu2}. $C_{II}$ is an overall factor
which will be  determined in this section.

There are 10 possible ways for the dividing degeneration limit
(one is $a_2-a_1=u$, $a_3-a_1=vu$ and $u\to 0$)\footnote{There are
some fine points which should be taken into account. See Sect. 3.2
of \cite{Phong3} for details.} and by using the result of
\cite{XiaoZhu} we have:
\begin{eqnarray}\label{4PointsFact}
 \mathcal{A}_{II}
& \to &
 C_{II} \,
 K(k_i, \epsilon_i) \, {10\over 6!} \,
 \, { { 2 \pi \over (\alpha')^3 } \over - s + { 4 \over \alpha'}
 }   \int {|K_1K_2/4|^2\over  T_1^5T_2^5}
 {{\rm d}^2 a_1\over|a_{14}a_{15}a_{16}|^2}
 {{\rm d}^2v\over|v(v-1)|^2} \nonumber \\
& & \times {{\rm d}^2x_1{\rm d}^2x_2\over|y_1(x_1)y_1(x_2)|^2}
 {{\rm d}^2z_3{\rm d}^2z_4\over|y_2(z_3)y_2(z_4)|^2}
  \nonumber \\
& &\times \exp\Big\{ -
 \Big( G_1(x_1,x_2)- G_1(x_1,p_1) - G_1(x_2,p_1)\Big)
 \nonumber \\
& & \qquad -\Big( G_2(z_3,z_4) - G_2(z_3,p_2) - G_2(z_4, p_2)
 \Big) \Big\}\, . \label{factorc}
\end{eqnarray}
The one-loop amplitude in hyperelliptic language is:
\begin{eqnarray}
& & \hskip -1cm {1\over 4!} \int {|K|^2 \over  T^5}
 {{\rm d}^2 a_1\over|a_{12}a_{13}a_{14}|^2}
  \times {{\rm d}^2z_1{\rm d}^2z_2\over|y(z_1)y(z_2)|^2}
   \nonumber \\
 & &\times \exp\Big\{ -
 \Big( G(z_1,z_2)- G(z_1,z_3) - G(z_2,z_3)\Big)
 \nonumber \\
 & = &  {1 \over  (2 \pi)^2 }
  \int_F { {\rm d}^2 \tau \over ({\rm Im}\tau)^5 } \, \int
\prod_{i=1}^2
{{\rm d}^2 z_i } \nonumber \\
& &  \times \prod_{   r<s}^3\left| {\Theta_1(z_{rs}|\tau) \over
\partial \Theta_1(0|\tau) } \, {\rm exp}\left(  {  -\pi \over {\rm
Im} \tau }({\rm Im}z_{rs})^2 \right) \right|^{\alpha' k_r \cdot
k_s } . %\label{eqfourty}
 \end{eqnarray}
This gives the following relation by using factorization relation:
\begin{equation}
C_{II} \, {10  \over 6!}   \times {  \pi \over 2^3 (\alpha')^3 }
\, \times \left( 4! \over (2 \pi)^2 \right)^2  = ( g_3^{\rm
1-loop} )^2 .
\end{equation}
This gives
\begin{equation}
C_{II} = {2^4 \, g_c^6 \over (\alpha')^7 \pi}.
\end{equation}
In period matrix language we have\footnote{The factor $\pi\over 2$
in the last equation of \cite{WuZhu2} should be $2\pi$.}:
\begin{eqnarray}
\mathcal{A}_{II} & =  &
 C_{II} \,{ 1 \over (2\pi)^6 \,
2^5} \,
 K(k_i, \epsilon_i) \, \int {|{\rm d}^3 \tau|^2 \over
 (\det {\rm Im})^5 } \nonumber \\
 & & \times  \int \prod_{i=1}^4 {\rm d}^2 z_i
 |3 {\cal Y}_s|^2 \, \prod_{i<j}
 \exp\{-k_i\cdot k_j \langle X(z_i)X(z_j) \rangle
\} ,
 \end{eqnarray}
and the overall coefficient is
\begin{equation}
\hat C_{II} =  C_{II}  \, { 1 \over (2\pi)^6 \, 2^5}  = {g_c^6
\over (2 \pi \alpha')^7} .
\end{equation}
This result agrees with D'Hoker, Gutperle and Phong \cite{Phong3}
by taking into account the different convention for ${\rm d}^2 z$
(we use ${\rm d}^2 z= {\rm d} x {\rm d}y$ for $z = x+ i y$).

\section*{Appendix A: Formulas for tensor integration}

Here is a list of all the formulas needed for tensor integrations
which are used in Appendix B to prove eq.~(\ref{eqappendixc}). We
have
\begin{equation} \int { {\rm d}^D p \over (2 \pi )^D} \, p^\mu
\delta(p^2) \, \delta( (p+k)^2) = - { 1\over 2} \, k^\mu \int \, {
{\rm d}^D p \over (2 \pi )^D} \, \delta(p^2) \, \delta( (p+k)^2) .
\end{equation}

\begin{eqnarray}
&&\hspace{-1cm}\int { {\rm d}^D p \over (2 \pi )^D} \,
p^{\mu_1}p^{\mu_2}
\delta(p^2) \, \delta( (p+k)^2)
\nonumber\\
& = & {1\over 4(D-1)}(-k^2\,\eta^{\mu_1\mu_2} + D \,
k^{\mu_1}k^{\mu_2}) \nonumber \\
& & \times \int \, { {\rm d}^D p \over (2 \pi )^D} \, \delta(p^2)
\, \delta( (p+k)^2) .
\end{eqnarray}
\begin{eqnarray}
&&\hspace{-1cm}\int { {\rm d}^D p \over (2 \pi )^D} \,
p^{\mu_1}p^{\mu_2}p^{\mu_3}p^{\mu_4}
\delta(p^2) \, \delta( (p+k)^2)
\nonumber\\
&=&{1\over 16(D^2-1)}\Big(k^4\big(\eta^{\mu_1\mu_2}
\eta^{\mu_3\mu_4} +(\mu_1\leftrightarrow
\mu_3)+(\mu_1\leftrightarrow \mu_4)\big) \nonumber \\ & & -k^2
(D+2)\big(
 k^{\mu_1}k^{\mu_2}g^{\mu_3\mu_4}
+(\mbox{5 more terms}) \big) \nonumber \\ &&
+(D+2)(D+4)k^{\mu_1}k^{\mu_2}k^{\mu_3}k^{\mu_4}\Big) \nonumber\\&&
\times\int \, { {\rm d}^D p \over (2 \pi )^D} \, \delta(p^2) \,
\delta( (p+k)^2) .
\end{eqnarray}
\begin{eqnarray}
&&\hspace{-1cm}\int { {\rm d}^D p \over (2 \pi )^D} \,
p^{\mu_1}p^{\mu_2}p^{\mu_3}p^{\mu_4} p^{\mu_5}p^{\mu_6}
\delta(p^2) \, \delta( (p+k)^2)
\nonumber\\
&=&{1\over
64(D+3)(D^2-1)}\Big(-k^6\big(\eta^{\mu_1\mu_2}\eta^{\mu_3\mu_4}
\eta^{\mu_5\mu_6} +  (\mbox{14 more  terms})\big) \nonumber\\&&
+k^4 (D+4)\big(
 k^{\mu_1}k^{\mu_2} \eta^{\mu_3\mu_4}\eta^{\mu_5\mu_6}
+ (\mbox{ 44 more terms})\big) \nonumber \\ && -k^2
(D+4)(D+6)\big(k^{\mu_1}k^{\mu_2} k^{\mu_3}k^{\mu_4}
\eta^{\mu_5\mu_6}+  (\mbox{14 more terms})\big) \nonumber \\ && +
(D+4)(D+6)(D+8)k^{\mu_1}k^{\mu_2}k^{\mu_3}
k^{\mu_4}k^{\mu_5}k^{\mu_6} \Big) \nonumber\\&& \times\int \, {
{\rm d}^D p \over (2 \pi )^D} \, \delta(p^2) \, \delta( (p+k)^2) .
\end{eqnarray}

\begin{eqnarray}
& &\hspace{-1cm}\int { {\rm d}^D p \over (2 \pi )^D} \,
p^{\mu_1}p^{\mu_2}p^{\mu_3}p^{\mu_4} p^{\mu_5}p^{\mu_6}
p^{\mu_7}p^{\mu_8} \delta(p^2) \, \delta( (p+k)^2)
\nonumber\\
&=&{1\over 256(D+5)(D+3)(D^2-1)} \nonumber \\&&
\times\Big(k^8\big(\eta^{\mu_1\mu_2}\eta^{\mu_3\mu_4}
\eta^{\mu_5\mu_6}\eta^{\mu_7\mu_8} +  (\mbox{104 more terms})\big)
\nonumber \\ && -k^6 (D+6)\big(
 k^{\mu_1}k^{\mu_2}\eta^{\mu_3\mu_4}\eta^{\mu_5\mu_6}
 \eta^{\mu_7\mu_8}
+ (\mbox{ 419 more terms})\big) \nonumber \\&& +k^4
(D+6)(D+8)\big(
 k^{\mu_1}k^{\mu_2}k^{\mu_3}k^{\mu_4}\eta^{\mu_5\mu_6}
 \eta^{\mu_7\mu_8} + (\mbox{ 209 more })\big) \nonumber \\
&& -k^2 (D+6)(D+8)(D+10) \nonumber \\
& & \times \big(k^{\mu_1} k^{\mu_2}k^{\mu_3}k^{\mu_4}k^{\mu_5}
k^{\mu_6} \eta^{\mu_7\mu_8} + (\mbox{27 more })\big) \nonumber \\
&& + (D+6)(D+8)(D+10)(D+12)k^{\mu_1}k^{\mu_2}k^{\mu_3}k^{\mu_4}
k^{\mu_5}k^{\mu_6}k^{\mu_7}k^{\mu_8} \Big) \nonumber\\&&
\times\int \, { {\rm d}^D p \over (2 \pi )^D} \, \delta(p^2) \,
\delta( (p+k)^2) .
\end{eqnarray}

\section*{Appendix B: Proof of eq.~(\ref{eqappendixc}):
the summation over the intermediate states in  one-loop  unitarity
relation}

As we explained in subsection 4.2, in order to use the 1-loop
unitarity relation to determine the coefficient $g_{_M}^{\rm
1-loop}$, we need eq.~(\ref{eqappendixc}) to do the summation over
all the intermediate states. This equation is only true when it is
inserted  in an integral where we can use the substitution rules
implied by the formulas for tensor integrations derived in the
last appendix. Here we will give some details for the summation
over all the intermediate states and sketch a proof of
eq.~(\ref{eqappendixc}).

The summation over the intermediate states can be done separately
for the left-moving part and the right-moving part. So we need the
formulas for the $NS$ and $R$ intermediate states. For $NS$
intermediate states, the kinematic factor $K_M$ is given in
eq.~(\ref{eqtwoone}) and we have:
\begin{eqnarray}
& & \hskip -1cm \sum_{\epsilon_1,\epsilon_2}
|K_M(p_1,\epsilon_1;p_2,\epsilon_2;k,\alpha,\sigma)|^2
  =
-36\alpha_{\mu_1\nu \rho }(k){\alpha_{\mu_2}}^{\nu\rho}(-k)
p_1^{\mu_1}p_1^{\mu_2}
 \nonumber \\
 & & + \sigma_{\mu_1{\nu_1}}(k)\sigma_{\mu_2{\nu_2}}(-k) \Big\{
 (D-2)p_1^{\mu_1}p_1^{\mu_2}p_1^{\nu_1}p_1^{\nu_2} \nonumber \\
  & &
 \hskip 1cm -4p_1^{\mu_1}p_1^{\mu_2}\eta^{\nu_1\nu_2}
 +\eta^{\mu_1\mu_2}\eta^{\nu_1\nu_2}
\Big\} , \label{eqsixsix}
\end{eqnarray}
by using the following summation formula:
\begin{equation}
\sum_\epsilon \epsilon^\mu(p)\epsilon^\nu(-p) =
\eta^{\mu\nu}-{p^{\mu}p'^{\nu}\over 2|p|^2} -{p'^{\mu}p^{\nu}\over
2|p|^2} ,
\end{equation}
where $p=(p^0,p^1,p^2,p^3)$ and $p'=(-p^0,p^1,p^2,p^3)$. $D$ is
the dimension of space-time.

For $R$ intermediate states, the kinematic factor $K_{MFF}$ is
given in eq.~(\ref{eqthreesix}) and we have:
\begin{eqnarray}
& & \hskip -1cm \sum_{u_1,u_2} |K_{MFF}(k,\alpha,\sigma;
p_1,u_1;p_2,u_2;)|^2 \nonumber \\
& = & \alpha_{\mu_1\nu_1\rho_1}(k) \alpha_{\mu_2\nu_2\rho_2}(-k)
\, A^{\mu_1\nu_1\rho_1
\mu_2\nu_2\rho_2}      \nonumber \\
 & &  -\frac{N}{2}\sigma_{\mu_1{\nu_1}}(k)
 \sigma_{\mu_2{\nu_2}}(-k)
\big(  p_1^{\mu_1}  p_1^{\mu_2}p_1^{\nu_1}p_1^{\nu_2}
 +\frac{k^2}{4} \eta^{\nu_1\nu_2} \, p_1^{\mu_1}p_1^{\mu_2}
\big)  , \label{eqsixeight}
\end{eqnarray}
where $N=32$ is the dimension of Dirac spinor and
\begin{eqnarray}
A^{\mu_1\nu_1\rho_1\mu_2\nu_2\rho_2}  & = & \frac{1}{8}
 {\rm Tr}\Big[  p\sl_1\Gamma^{[{\mu_1}{\nu_1}{\rho_1}]}
 p\sl_2\Gamma^{[{\mu_2}{\nu_2}{\rho_2}]}  {1+\Gamma^{11}\over 2}
    \Big] \nonumber \\
 & =& \frac{1}{16}\, {\rm Tr}
 \Big[p\sl_1\Gamma^{[{\mu_1}{\nu_1}{\rho_1}]}
 p\sl_2\Gamma^{[{\mu_2}{\nu_2}{\rho_2}]}  \Big] .
\end{eqnarray}
Here we have used:
\begin{equation}
\sum_{u} u(p)  \, \bar u(-p) = {1+\Gamma^{11}\over
2}p\sl.
\end{equation}

The 4 different contributions as displayed in Fig.~4 are given as
follows:
\begin{eqnarray}
{\cal A}_{M2(NS-\widetilde{NS})} & = & g_{_M} \,
K_M(k_1,\epsilon_1;
k_2,\epsilon_2; k, \alpha,\sigma) \nonumber \\
& & \times  K_M(k_1,\tilde\epsilon_1; k_2,\tilde\epsilon_2; k,
\tilde\alpha,\tilde\sigma)  , \\
{\cal A}_{M2(R-\tilde{R})} & =   &  - g_{_{MRR}} \,
K_{MFF}(k_1,u_1;
k_2,u_2; k, \alpha,\sigma) \nonumber \\
& & \times  K_{MFF}(k_1,\tilde u_1; k_2,\tilde u_2; k,
\tilde\alpha,\tilde\sigma)  , \\
{\cal A}_{M2(R-\widetilde{NS})} & = & g_{_{MFF}}\,
K_{MFF}(k_1,u_1;
k_2,u_2; k, \alpha,\sigma) \nonumber \\
& & \times  K_M(k_1,\tilde\epsilon_1; k_2,\tilde\epsilon_2; k,
\tilde\alpha,\tilde\sigma)  , \\
{\cal A}_{M2(NS-\tilde{R})} & = & g_{_{MFF}} \, K_{MFF}(k_1,u_1;
k_2,u_2; k, \alpha,\sigma) \nonumber \\
& & \times  K_M(k_1,\tilde\epsilon_1; k_2,\tilde\epsilon_2; k,
\tilde\alpha,\tilde\sigma)   .
\end{eqnarray}
By using these results we have
\begin{eqnarray} \sum_{\rm all~intermediate~states}
|{\cal A}_{M**}^{\rm tree}|^2 & = & \sum_{\epsilon_i,
\tilde\epsilon_i} |{\cal A}_{M2(NS-\widetilde{NS})}|^2 +
\sum_{u_i, \tilde u_i} |{\cal A}_{M2(R-\tilde{R})}|^2 \nonumber
\\
& & \hskip -2cm + \sum_{\epsilon_i, \tilde u_i} |{\cal
A}_{M2(NS-\tilde{R})}|^2 + \sum_{u_i, \tilde\epsilon_i} |{\cal
A}_{M2(R-\widetilde{NS})}|^2
\end{eqnarray}
Now we can use eqs.~(\ref{eqsixsix}) and (\ref{eqsixeight}) to do
the summation over the intermediates. The results can be
simplified further by using the formulas in Appendix A for the
integration over $k_1$ ($k_2 = - (k+k_1)$ by momentum
conservation). After a long and tedious calculation,
eq.~(\ref{eqappendixc}) is proved.

\section*{Acknowledgments}

C.-J. Zhu would like to thank Edi Gava, Roberto Iengo and K. S.
Narain for their interests in this work. He would also like to
thank Prof. S. Randjbar-Daemi  and the hospitality at Abdus Salam
International Centre for Theoretical Physics, Trieste, Italy.

\end{document}